\documentclass[journal]{IEEEtran}
\usepackage{cite}

\ifCLASSINFOpdf
  \usepackage[pdftex]{graphicx}
  \graphicspath{{../fig/}{../jpeg/}{./}}
  \DeclareGraphicsExtensions{.pdf,.jpeg,.png}
\else
  \usepackage[dvips]{graphicx}
  \graphicspath{{../eps/}}
  \DeclareGraphicsExtensions{.eps}
\fi
\usepackage{amsmath}
\usepackage{amsfonts}
\usepackage{amssymb}
\usepackage{multirow}
\usepackage{multicol}
\usepackage{color}

\usepackage{url}

\begin{document}
%
\title{Direct Speech-to-Image Translation}
%
%
%

\author{Jiguo Li,
        Xinfeng Zhang,~\IEEEmembership{Member,~IEEE,}
        Chuanmin, Jia,
        Jizheng Xu,~\IEEEmembership{Senior Member,~IEEE,}
        Li Zhang,
        Yue Wang,
        Siwei Ma,~\IEEEmembership{Member,~IEEE,}
        and~Wen Gao~\IEEEmembership{Fellow,~IEEE,}
\thanks{Jiguo Li is with the Key Lab of Intelligent Information Processing, Institute of Computing Technology, Chinese Academy of Sciences, Beijing 100190, China, the University of Chinese Academy of Sciences, Beijing 100049, China, and also with the National Engineering Laboratory for Video Technology, School of Electronic Engineering and Computer Science, Peking University, Beijing 100871, China (e-mail: jiguo.li@vipl.ict.ac.cn)}
\thanks{Xinfeng Zhang is with the School of Computer Science and Technology, University of Chinese Academy of Sciences, Beijing 100049, China (e-mail: xfzhang@ucas.ac.cn)}
\thanks{Siwei Ma, Chuanmin Jia, Wen Gao are with the Institute of Digital Media, School of Electronic Engineering and Computer Science, Peking University, Beijing 100871, China, and are also with the Peng Cheng Lab, Shenzhen, China (e-mail: {swma, cmjia, wgao}@pku.edu.cn).\textit{(Corresponding author: Prof. Siwei Ma)}}
\thanks{Jizheng Xu, Li Zhang, Yue Wang are with Bytedance Inc. (e-mail: {xujizheng, lizhang.idm, wangyue.v}@bytedance.com)}
}

%
%

\markboth{Journal of Selected Topic on Signal Processing,~Vol., No., January~2020}%
{Shell \MakeLowercase{\textit{et al.}}: Bare Demo of IEEEtran.cls for IEEE Journals}
%



\maketitle


\begin{abstract}
  Direct speech-to-image translation without text is an interesting and useful topic due to the potential applications in human-computer interaction, art creation, computer-aided design. etc. Not to mention that many languages have no writing form. However, as far as we know, it has not been well-studied how to translate the speech signals into images directly and how well they can be translated. 
  {In this paper, we attempt to translate the speech signals into the image signals without the transcription stage.} 
  Specifically, a speech encoder is designed to represent the input speech signals as an embedding feature, and it is trained with a pretrained image encoder using teacher-student learning to obtain better generalization ability on new classes. Subsequently, a stacked generative adversarial network is used to synthesize high-quality images conditioned on the embedding feature. Experimental results on both synthesized and real data show that our proposed method is effective to translate the raw speech signals into images without the middle text representation. Ablation study gives more insights about our method.

\end{abstract}


%
\IEEEpeerreviewmaketitle

\section{Introduction}
%
%
%
%
\IEEEPARstart{I}t has been widely accepted by cognitive science community that infants begin learning their native language not by learning words, but by discovering the correlations between the speech signal and visual information~\cite{bergelson20126}. 
Infants know some aspects of their language by 6–12 months, while they do not understand the common native-language words until 12 months~\cite{thomas1981semantic}. 
When communicating with their parents, the infants only receive continuous speech signals from the parents and the visual signals from the surrounding. And the infants can learn the correlation between the high-frequency speech words and the objects or local visual textures. 
Thus, it is interesting to explore whether a machine can translate the speech signals into images directly, without the help of language words. 
{Translating data between different modalities is a cutting-edge area recently. However, speech-to-image translation has not been well-studied while the similar topic, text-to-image translation, have been investigated in recent literature~\cite{stackgan, stackgan++, attngan}.
Besides, many languages have no writing form, which calls for the approaches to understand and visualize the speech directly~\cite{tannen1982spoken}. Not to mention the potential applications in human-computer interaction, art creation and computer-aided design, where speech is the nature input and middle text representation is not necessary. So exploring speech-to-image translation is necessary and meaningful.}

\begin{figure}[t]
  \begin{center}
     \includegraphics[width=0.9\linewidth]{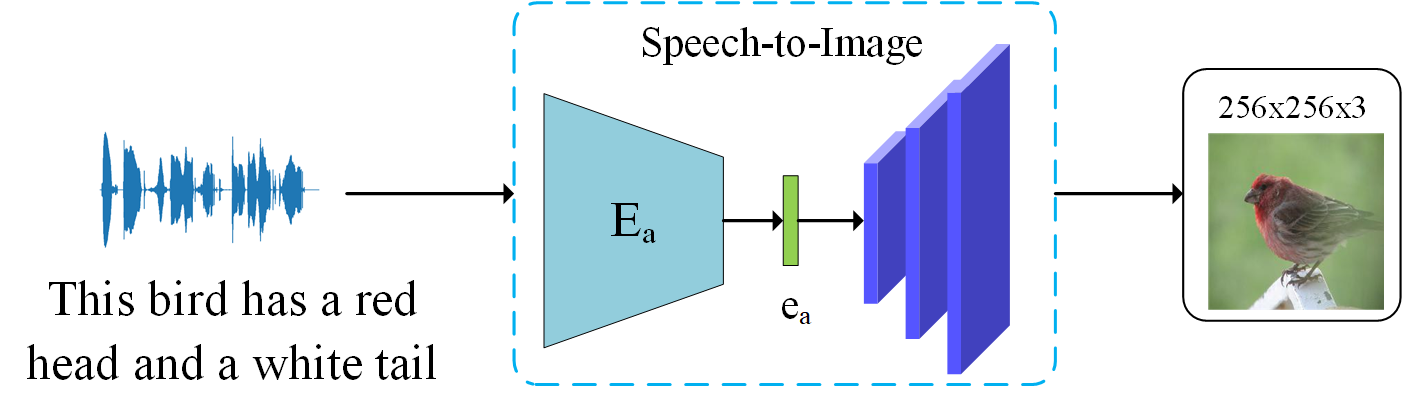}
  \end{center}
     \caption{Illustration for our task: speech-to-image translation without text. \textbf{Note: the text is shown only for readability, it is not used in the speech-to-image model.}}
  \label{fig:audio2image}
  \end{figure}

As illustrated in Fig.~\ref{fig:audio2image}, given the raw speech descriptions: ``this bird has a red head and a white tail", the corresponding images can be synthesized, which means that the machine has understood the speech signal to some extent and been able to translate the semantic information in the speech signal into the image. 
  {Speech and image are in different modalities and the modality gap between these two types of data makes direct speech-to-image translation not trivial. Text-to-image translation~\cite{reed2016learning, stackgan, stackgan++, attngan} is a closely related topic to ours, which has been investigated for several years. In some text-to-image models, zero-shot learning based methods~\cite{reed2016learning, oh2019speech2face} and generative adversarial networks~(GANs)~\cite{gan} have been used to extract features and synthesize realistic images, respectively. These models generalize better on the new testing classes by leveraging the teacher-student learning to train the text encoder~\cite{reed2016learning, reed2016generative}. Compared with the text-to-image translation, speech-to-image translation might be more challenging because the speech signals are continuous, unaligned and noisy.} 
In addition to the text-to-image translation, several models for audio-to-image generation were also presented in recent years. Chen~\textit{et al.}~\cite{chen2017deep} and Hao~\textit{et al.}~\cite{hao2018cmcgan} synthesized instrument images from different music inputs; 
Oh~\textit{et al.}~\cite{speech2face} and Amanda~\textit{et al.}~\cite{wav2pix2019icassp} reconstructed the human face images from input speech based on the positive correlations between a person’s appearance and his voice, and both their frameworks contain a speech encoder and a face decoder.
Different from these audio-to-image generation works, which model the acoustic or phonetic information mainly, our speech-to-image translation aims to model the linguistic information in the input speech and translate it into the images.
Recent works about audio-visual correlation learning~\cite{harwath2017learning, harwath2018jointly} have shown it is feasible to learn the correlation between visual speech descriptions and the objects or local texture in the images, forming the basis of our speech-to-image translation task.
In some other topics about speech processing, such as speech-to-speech translation and speech keyword search, recent works~\cite{jia2019direct, Audhkhasi2017end} have attemped to translate or search the speech without the help of the transcription text, indicating it is feasible to understand the linguistic information in the speech without the help of the middle text representations.

Highly inspired by these previous related works, we design a model to extract features from speech data and train the model in the teacher-student learning manner. 
In particular, the speech signal is firstly represented as a low-dimensional embedding feature via a speech encoder, then this feature is fed into a conditional generative adversarial network as the condition, and the generator synthesizes the corresponding image with semantic consistency. 
To the best of our knowledge, our work is the first one to attempt to translate the speech signals into images without the help of text. 
Compared with the straightforward ``two-stage'' method, the classifier-based method and the text-to-image models, our method shows better performance than the ``two-stage'' method and the classifier-based method, even achieves comparable performance to the text-to-image models on the synthesized datasets. Experiments on the real speech data also show the potential for the real application scenarios, like human-computer interaction. etc.

The main contributions of this work can be summarized as follows:
\begin{itemize}
    \item We propose a framework to translate the speech signals into images directly. Experiments on the synthesized data and real data demonstrate the effectiveness of our proposed framework.

    \item We train the speech encoder via teacher-student learning that transfers the knowledge in a pretrained image encoder into the speech encoder. Experiments on the synthesized data show that our method can learn the semantic information in the speech descriptions better than the previous classifier-based method, and provide better translation results.
    
    \item Ablation study about the loss items, image scales and feature interpolation gives more insights about our method and the speech-to-image translation problem.
\end{itemize}

The rest of this paper is organized as follows: Section~\ref{sec:related_works} briefly reviews related works on generative adversarial networks, text-to-image translation, audio-to-image generation, audio-visual correlation learning, teacher-student learning, and direct speech translation, Section~\ref{sec:proposed_model} presents our speech-to-image model in detail, and Section~\ref{sec:experiments} introduces and analyzes  the experimental results on both synthesized and real data, Section~\ref{sec:ablation_study} conducts the ablation study. Finally, Section~\ref{sec:conclusion} concludes this paper.

\section{Related Works}\label{sec:related_works}
In this section, we review the related works on generative adversarial networks, text-to-image translation, audio-to-image generation, audio-visual correlation learning, teacher-student learning, and direct speech translation.

\subsection{Generative Adversarial Networks}
Generative adversarial networks~(GANs) have drawn much attention since it was presented by Goodfellow~\textit{et al.}~\cite{gan} due to its ability to generate high-dimensional data,~\textit{e.g.} images. In the model, the generators aim to generate fake data that cannot be separated from the real data, while the discriminators aim to differentiate the generated fake data from the real data. The whole model is optimized via the following loss functions~\cite{gan, stackgan++}:
\begin{align}
\min_G\max_D~\mathbb{E}_{x\sim p_{data}}[\log{D(x)}]+\mathbb{E}_{z\sim p_z}[\log{(1-D(G(z)))}],
\end{align}
where $G$ is the generator and $D$ is the discriminator.
It is a two-player zero-sum game to arrive a local Nash equilibrium~\cite{heusel2017gans}, at which neither the discriminator nor the generator can decrease its respective loss. The generator learns a mapping between the noise distribution~(\textit{e.g.} the uniform or Gaussian) and the real data~(\textit{e.g.} the images or text). 
When synthesizing images via GANs, attributions~\cite{huang2017beyond}, text descriptions~\cite{reed2016generative}, sketches~\cite{isola2017image}, or images with another style~\cite{zhu2017unpaired} have been used as the conditions to control the appearance of the generated images. 

\subsection{Text-to-Image Translation}
Text-to-image translation aims to synthesize images which are semantically consistent with the input text descriptions. It is challenging due to the modality gap between text and images. 
The computer vision and machine learning community did not pay much attention to this challenging problem until Reed \textit{et al.}~\cite{reed2016generative} used a GAN to synthesize the images conditioned on a low-dimensional representation extracted from the text description. Following Reed's work~\cite{reed2016generative}, StackGAN~\cite{stackgan} and StackGAN V2~\cite{stackgan++} were proposed to generate photo-realistic images up to a resolution of $256\times256$ from the text descriptions via a pretrained text encoder~\cite{reed2016learning}. Multi-scale discriminators for increasing resolutions were used in StackGAN and StackGAN V2 to generate images progressively because synthesizing images with a high resolution in one stage had been demonstrated with difficulty~\cite{harwath2015deep}. 
Besides, spatial attention was applied in text-to-image translation~\cite{attngan} by training a multimodal similarity model to calculate the similarity between the word embedding features and the local image features. With text encoder trained by teacher-student learning, these text-to-image translation models generalize well on the new testing classes.

\subsection{Audio-to-Image Generation}
Based on the correlation between audio and images, such as music and instruments, human voices and face appearances,  audio-to-image generation aims to generate the images paired with the input audio signals. 
Chen~\textit{et al.}~\cite{chen2017deep} firstly attempted to generate instrument images from the music by leveraging a classifier-based feature extractor and a GAN, but another model is needed if we want to generate music from the image. Hao~\textit{et al.}~\cite{hao2018cmcgan} proposed a uniform framework using cycle constraint for the visual-audio mutual generation. Recently, Oh~\textit{et al.}~\cite{oh2019speech2face} presented a model for generating the face images from a voice using a pretrained face decoder, but the generated results are not sharp due to the concern of privacy. Duarte~\textit{et al.}~\cite{wav2pix2019icassp} generated sharp face images conditioned on the input speech segmentation using GANs.
Different from these previous audio-visual generation works, speech-to-image translation aims to capture the linguistic information in the speech signals and generate images semantically consistent with the input speech descriptions.

\subsection{Audio-Visual Correlation Learning}
Audio-visual correlation learning aims to learn a joint embedding feature space over both audio (\textit{e.g.} music, speech, nature sound,~\textit{etc.}) and visual (\textit{e.g.} images, videos,~\textit{etc.}) data using an embedding alignment model. Based on the prior works on text embedding~\cite{socher2010connecting}, Harward~\textit{et al.}~\cite{harwath2015deep} firstly investigated this task to align visual objects and speech signals by a region convolutional neural network~(RCNN)~\cite{girshick2014rich} and a spectrogram convolutional neural network~\cite{bengio2014word}. Furthermore, Harward~\textit{et al.}~\cite{harwath2018vision} used vision as an interlingual semantic embeddings of unaligned audios without the use of linguistic transcriptions or conventional speech recognition technologies. Recently, Harward~\textit{et al.}~\cite{harwath2017learning, harwath2018jointly} operated directly on the image pixels and speech waveforms to associate segments of visual audio captions without relying on any labels, segmentation information or alignment between these modalities. 
In addition to learning an embedding feature space for speech and images, our work further generates images from the speech embeddings.

\subsection{Teacher-Student Learning}
Teacher-student learning is a transfer learning approach, where a pretrained teacher model is used to “teach” a student model~\cite{manohar2018teacher}. It is widely used in model compression~\cite{ba2014deep, hinton2015distilling} and domain adaption~\cite{yu2013kl}. Reed~\textit{et al.}~\cite{reed2016learning} firstly used the GoogLeNet~\cite{googLeNet} pretrained on ImageNet~\cite{deng2009imagenet} as the teacher network to learn the deep representation for zero-shot tasks. Following this work, several text-to-image models~\cite{reed2016generative, stackgan, stackgan++} were proposed for the text-to-image task, based on the same teacher-student learning method to train the text encoder. 
Recently, teacher-student learning was used to generate the face behind a voice~\cite{speech2face}. As a comparison, traditional audio-to-image generation models~\cite{chen2017deep} used a classifier as the feature extractor. In our experiments, we compare the teacher-student learning method with the classifier-based methods and show that the teacher-student learning performs better.

\subsection{Direct Speech Translation}
Speech-to-speech translation is one of the most challenging tasks in speech processing and machine learning community, which has tremendous applicable value in our daily life. A speech translation system typically has three components: automatic speech recognition~(ASR), machine translation~(MT) and text to speech synthesizer~(TTS)~\cite{waibel2015enhanced}. Using text as a middle representation to divide this difficult task into three stages have been used for several decades. 
Recently, with the development of deep learning~\cite{lecun2015deep}, which shows great potential to model complicated data distribution, researchers have attempted to solve the challenging speech translation without the middle text representation.
B{\'e}rard~\textit{et al.}~\cite{berard2016listen} firstly attempted to build an end-to-end speech-to-text translation system without the text transcription, and showed comparable performance on the synthesized data. 
Subsequently, Duong~\textit{et al.}~\cite{duong2016attentional} introduced an attentional model for speech-to-speech translation without speech-to-text transcription, showing the superiority on the low-resources languages. 
Recently, Jia~\textit{et al.}~\cite{jia2019direct} proposed a sequence-to-sequence model to directly translate the speech into another speech, showing comparable performance (only slightly underperform) to a baseline, which cascades of a direct speech-to-text translation model and a text-to-speech synthesis model, on two Spanish-to-English speech translation datasets. 
These results demonstrated that extracting semantic information from raw speech signals without the middle text representation is practical. 

Motivated by the different principles for understanding speech signals, we design a framework to translate the speech signals into images directly, without the help of middle text representation. 
Specifically, a speech encoder is designed to encode the raw speech signals into a low-dimensional embedding feature. The speech encoder is trained by the teacher-student learning manner via a pretrained image encoder. Subsequently, the speech embedding features are fed into a generator to synthesize images with semantic consistency. Experimental results on both synthesized and real data demonstrate that our proposed model is capable of translating speech into images without the help of middle text representation.

\begin{figure*}[ht!]
  \begin{center}
     \includegraphics[width=0.9\linewidth]{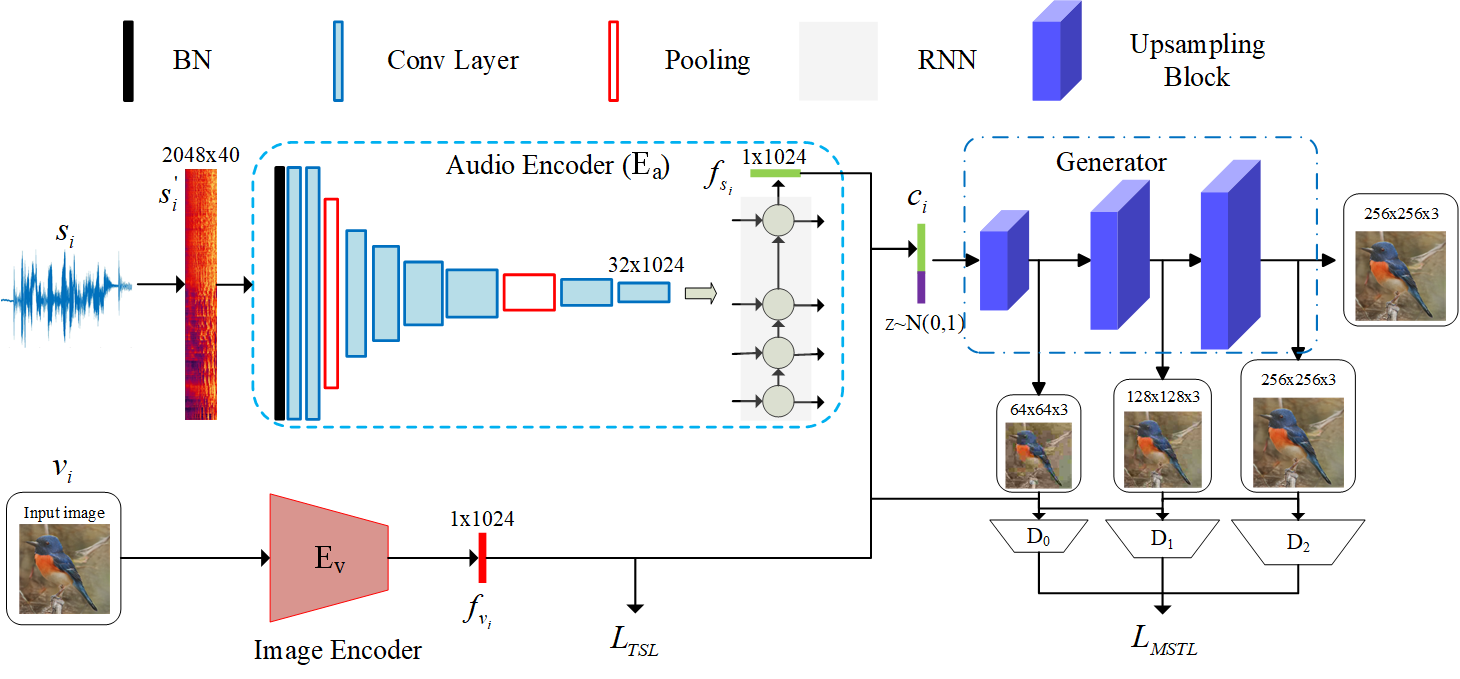}
  \end{center}
     \caption{Illustration for our speech-to-image model: a speech encoder, trained via teacher-student learning, is designed to extract a feature representation from the input speech. After encoding the raw speech signals into a low-dimensional embedding, our model synthesizes images at a resolution of $256\times256$ with semantic consistency from the embedding feature. In the figure, $TSL$ denotes teacher-student loss for the speech encoder and $MSTL$ denotes multi-scale triple loss for the generators.}
  \label{fig:proposal_model}
\end{figure*}

\section{Speech-to-Image Model}\label{sec:proposed_model}
The modality gap between speech signals and image signals makes it not feasible to directly regress the pixel value from speech signals. 
Inspired by the common text-to-image architectures~\cite{reed2016generative, stackgan, stackgan++, attngan}, we design a speech encoder to encode the speech signals into a low-dimensional embedding feature. Then this embedding feature is used to synthesize the corresponding images with semantic consistency, The diagram of the proposed algorithm is illustrated in Fig.~\ref{fig:proposal_model}. 


\subsection{Speech Encoder}
The input raw speech signal is first represented as a time-frequency spectrogram, then it is encoded by our speech encoder into a low-dimensional embedding feature with convolutional neural networks~(CNNs) and recurrent neural networks~(RNNs). Typically, the speech spectrogram is modeled via an RNN in ASR\cite{graves2013speech, hannun2014deep, amodei2016deep}, however, the long input spectrogram of our model may not convinent enough for RNN model optimization~\cite{hochreiter2001gradient}. Inspired by the character-based text embedding architecture~\cite{reed2016learning} and audio-visual cross-modal embedding learning~\cite{harwath2018jointly}, a multi-layer CNN is inserted before the RNN to reduce the signal length, as shown in Fig.~\ref{fig:proposal_model}. Given a speech signal $s_i$ and its spectrogram $s'_i$, the speech embedding $f_{s_i}$ can be obtained:
\begin{align}
  f_{s_i} = E_s(s'_i) = \text{RNN}(\text{CNN}(s'_i)),
\end{align}
where $E_s$ denotes our speech encoder. The input time-frequency spectrogram is firstly normalized along with frequency in the first layer of the CNN, then the output of the CNN is subsequently encoded as a 1024 dimensional embedding feature by an RNN. The input length of the RNN can be variable, with the only constraint that the input length of the CNNs should be longer than 64 because the model shortens the input sequence by 64 times.

\subsection{Teacher-Student Learning}
The optimization for our speech encoder is not trivial because the ground truth of the speech embedding is not available.
{Although the class label of the speech description is accessible, there might be generalization problem when the trained model is tested on the new unseen data (new class respect to the training set). Inspired by the cross-modal generation models~\cite{reed2016learning, stackgan, harwath2015deep, harwath2018jointly, speech2face}, in this work, we use teacher-student learning~\cite{hinton2015distilling} to overcome this problem to some extent.}

\indent Given an image and its speech description, $(v_i, s_i)$, an image encoder and a speech encoder are used to represent the image/speech as a low-dimensional embedding feature, respectively:
\begin{align}
    f_{v_i} &= E_v(v_i) \\
    f_{s_i} &= E_s(s'_i),
\end{align}
where $s'_i$ is the time-frequency spectrogram of $s_i$. $E_v$/$E_s$ denotes the image/speech encoder, $f_{v_i}$/$f_{s_i}$ is the low-dimensional embedding feature of the input image/speech. It is worth mentioning that the image encoder is pretrained on a large dataset, such as ImageNet~\cite{deng2009imagenet}, and it is fixed when training the speech encoder. In our model, the pretrained image encoder is the ``teacher", while the speech encoder is the ``student". 
The goal of teacher-student learning is to optimize the student model to learn a similar feature space with the teacher model. So the dimension of the student model's feature space should be the same as the teacher's. In our model, GoogLeNet~\cite{googLeNet} is used as the teacher model to represent the input image with a resolution of $256\times256$ as a 1024 dimensional feature. As a result, our speech encoder also encodes the input time-frequency spectrogram into a feature with 1024 dimensions.
\subsection{Generative Network}
The generative network is used to synthesize images conditioned on the speech embedding feature. Following the recent works about text-to-images~\cite{stackgan, stackgan++, attngan}, we use a stacked conditional GAN, also known as StackGAN v2~\cite{stackgan++}, to synthesize the images due to its promising performance on generating photo-realistic images. 
As illustrated in Fig.~\ref{fig:proposal_model}, three branches are used in the generator to synthesize images with a resolution of $256\times256$, and three discriminators are used to distinguish the generated images with a resolution of $64\times64$, $128\times128$, $256\times256$ from the real images, respectively. Each upsampling block in the generator contains an upsampling layer and two residual blocks~\cite{he2016deep} to synthesize details based on the input low-resolution image. Given the condition $c_i$ and the input noise $z_i$ which samples from the Gaussian distribution, the generator synthesizes the fake images:
\begin{align}
  v_{if} = G(c_i, z_i),
\end{align}
where $v_{if}$ denotes the synthesized fake images, $G$ denotes the stacked generative network. In our model, the embedding feature of the input speech~($f_{s_i}$ in Fig.~\ref{fig:proposal_model})~is used as the condition $c_i$ for the generator.

\subsection{Training}
The whole model is trained with two steps. Firstly, the speech encoder is optimized with the image encoder. Secondly, the generator is trained conditioned on the speech embedding representations, which are extracted by the pretrained speech encoder. 

\subsubsection{Training the Speech Encoder} The speech encoder is trained via teacher-student learning~\cite{hinton2015distilling}, which transfers knowledge from a large model into a small model. In our model, the knowledge in the pretrained image encoder needs to be transferred into our speech encoder. We can optimize the norm to train the speech encoder, however, training student network with the norm optimization alone is slow and unstable~\cite{speech2face}. To stabilize and accelerate the training, additional loss items are introduced. Taking inspiration from the text-image embedding learning~\cite{reed2016learning} and audio-to-image generation~\cite{speech2face} models, norm loss, jointly embedding loss~(JEL), and knowledge distilling loss~(KDL) are used in our teacher-student loss~(TSL) for training the speech encoder. Given a batch of triplet data $(v_i, s_i, y_i)$, the spectrogram $s'_i$, the image encoder $E_v$ and the speech encoder $E_s$, the objective function between image and speech encoder is defined as:
\begin{align}\label{eq:se_loss}
  \mathcal{L}_{TSL_i} &= \mathcal{L}_{JEL_i} +\lambda_{norm} \mathcal{L}_{norm_i} + \lambda_{KDL} \mathcal{L}_{KDL_i} \\
  \mathcal{L}_{JEL_i} &= \alpha \mathbb{E}_{y_j \not= y_i}[\max(0, f_{s_i}^T f_{v_j} - f_{s_i}^T f_{v_i} + m_{diff})] \nonumber \\
   & + \beta \mathbb{E}_{y_j = y_i} [\max(0, f_{s_i}^T f_{v_j} - f_{s_i}^T f_{v_i} + m_{same})] \\
  \mathcal{L}_{norm_i} &=  \Sigma_k | f_{s_{ik}} - f_{v_{ik}} | \\
  \mathcal{L}_{KDL_i} &= \mathbb{KL}(softmax(f_{s_i})|softmax(f_{v_i})),
\end{align}
where $y_i$ is the class label for $v_i$ and $s_i$, $\lambda_{norm}, \lambda_{KDL}$ are hyper-parameters for fusing the three items. Following~\cite{speech2face}, $\lambda_{norm}$ and $\lambda_{KDL}$ are tuned to make the gradient magnitudes of the three items with respect to $f_{s_i}$ be with a similar scale at an early training iteration. In $\mathcal{L}_{JEL}$, $m_{diff}$/$m_{same}$ is the margin for $(v_i, s_i)$ pairs with different/same class label in a batch data, respectively. $\alpha$/$\beta$ is set to control the inter/intra class distance, respectively. Here, we optimize the 1-norm in $\mathcal{L}_{norm}$. $f_{s_i}=E_s(s'_i)$ is the embedding feature of the input speech description. $f_{v_i}=E_v(v_i)$ is the embedding represention of the input image. In $\mathcal{L}_{KDL}$, $softmax(x_i)=e^{x_i}/\Sigma_j e^{x_j}$. 

\subsubsection{Training the Generator} Following~\cite{stackgan++}, the generator is trained by a multi-scale triplet loss~(MSTL), which includes three items in each discriminator's loss: conditional item, unconditional item and wrong pair item. Given a triplet $(c_i, v_i, v_i^w)$, where $v_i^w$ denotes the wrong image which belongs to the different classes with $v_i$ :
\begin{align}
  \mathcal{L}_{D} &= \Sigma_{s=1}^{3} (\mathcal{L}_{D_s}^{cond}+\mathcal{L}_{D_s}^{uncond}+\mathcal{L}_{D_s}^{wrong}) \\
  \mathcal{L}_{D_s}^{cond} &= \mathbb{E}_{(v_i,c_i) \sim p} [ D_s(v_i, c_i)+(1-D_s(G(z, c_i), c_i)) ] \\
  \mathcal{L}_{D_s}^{uncond} &= \mathbb{E}_{(v_i,c_i) \sim p} [D_s(v_i)+(1-D_s(G(z, c_i)))] \\
  \mathcal{L}_{D_s}^{wrong} &= \mathbb{E}_{ (v_i,c_i) \sim p} [(1-D_s(v_i^w, c_i))] \\
  \mathcal{L}_{G} &= \Sigma_{s=1}^{3} \mathbb{E}_{(v_i,c_i)\sim p} [ D_s(G(z,c_i))+D_s(G(z,c_i),c_i) ],
\end{align}
where $p$ is the data distribution for the pair $(v_i, c_i)$, $\mathcal{L}_D$ and $\mathcal{L}_G$ both contain three scales. The loss for each discriminator has three items to model both conditional and unconditional distribution. By contrast, the traditional conditional GAN~\cite{cgan} only contains the conditional item $\mathcal{L}_{D_s}^{cond}$, without taking the unconditional distribution into account.

\subsection{Inference}
In the inference phase, the time-frequency spectrograms of the input speech descriptions are encoded as low-dimensional embedding features by the speech encoder. Subsequently, the generator synthesizes images with a resolution of $256\times256$ conditioned on the embedding features. The inference can be denoted as:
\begin{align}
  v_{if} = G(z, E_s(s'_i)),
\end{align}
where $z$ is the noise vector, $s'_i$ is the spectrogram of the input speech, $v_{if}$ is the synthesized images semantically consistent with the input speech descriptions.


\subsection{Implementation Details}
The raw speech signals are represented as log Mel filter bank spectrograms, following~\cite{harwath2018jointly}. Specifically, the DC component of each audio is removed via mean subtraction, followed by the pre-emphasis filtering and the short-time Fourier transform (STFT) computation by using a 25 ms Hamming window with 10 ms shift. Then the squared magnitude spectrum of each frame is taken into consideration and the log energies with each of 40 Mel filter bands are computed. 
As a result, the spectrograms with shape $(\text{band}, \text{frame})$, where the band here is $40$ with variable frame number, can be obtained. 
When training the speech encoder, we use GoogLeNet~\cite{googLeNet} as the image encoder. As for the hyperparameters, we set $\lambda_{norm}=5, \lambda_{KDL}=1000, m_{diff}=1, m_{same}=0.1$ after tuning them to balance the gradients respect to the speech embedding feature. 

\section{Experiments Results and Ayalyses}\label{sec:experiments}
In this section, we verify the effectiveness of the proposed model for translating speech signals into images without middle text representations and show how well our model can achieve. Firstly, we generate images from the synthesized speech data to demonstrate the effectiveness of our model and compare our model with the straightforward ``two-stage'' method, traditional classifier-based method, and text-to-image models. Secondly, experiments on real data are conducted to explore our model's robustness to the real noise and the potential for the real application scenarios. Finally, ablation study about the different loss items,  image scales, and feature interpolation gives more insights about our model. 

\subsection{Datasets and Metrics}
\textbf{Datasets.} Several datasets, like Places 205 dataset~\cite{zhou2014learning}, Caltech-UCSD Birds 200-2011~(CUB-200)~\cite{WahCUB_200_2011},  Oxford Flower with 102 categories~(Oxford-102)~\cite{Flower} and Microsoft COCO~\cite{COCO}, are used in the audio-visual correlation learning or text-to-image translation. Harwath~\textit{et al.} \cite{harwath2018jointly} used the Places 205 dataset with speech descriptions~\cite{harwath2017learning, harwath2016unsupervised} to learn the association between spoken audio caption segments and nature images portions. 
Reed \textit{et al.}~\cite{reed2016learning} used CUB-200 to learn deep representations of visual text descriptions. Most investigations about text-to-image translation~\cite{reed2016generative, stackgan, stackgan++} conducted their experiments on CUB-200, Oxford-102, or COCO dataset. 
However, these datasets only contain the text descriptions and do not have any speech label, making us difficult to leverage the dataset off-the-shelf to conduct the experiments. Fortunately, with the development of text-to-speech generation~\cite{deep_voice2, deep_voice, deep_voice3} technologies based on deep learning and large-scale labeled speech datasets~\cite{hernandez2018ted, panayotov2015librispeech}, we can use some mature commercial text-to-speech systems like Baidu TTS eigine\footnote{https://cloud.baidu.com/product/speech/tts} or Microsoft Bing TTS\footnote{https://azure.microsoft.com/en-us/services/cognitive-services/text-to-speech/} to synthesize large-scale high-quality speech from the text descriptions. 
The synthesized speech data are continuous and unaligned, just like real speech, although they have no background noise and speaker variance.

\indent To compare our model with the text-based models, we use CUB-200~\cite{WahCUB_200_2011} and Oxford-102~\cite{Flower} dataset to test the performance of our model and synthesize the speech descriptions from the text descriptions via Baidu TTS. 
As illustrated in Table~\ref{tab:dataset_info},CUB-200~\cite{WahCUB_200_2011} dataset contains 11788 bird images classified into 200 categories, with one bounding box and 10 sentences text descriptions per image. Oxford-102~\cite{Flower} dataset has 8189 flower images classified into 102 classes.
The speech descriptions are synthesized by Baidu TTS engine from the text captions in both datasets. Following~\cite{stackgan++}, we crop all the images to ensure that bounding boxes have greater-than-0.75 object-image-size ratios for CUB-200 dataset. We also conduct experiments on a subset of Places 205 dataset~\cite{harwath2017learning, harwath2016unsupervised} with real speech descriptions to explore the potential for real applications. 

\textbf{Evaluation Metrics.}  Generally, it is difficult to fairly evaluate the generative models. As in~\cite{reed2016generative, stackgan, stackgan++}, we use inception score~\cite{inception_score} and Fr\'echet inception distance~\cite{dowson1982frechet, fid} to quantitatively evaluate our models. They are formulated as, 
\begin{align}
    IS &= \exp(\mathbb{E}_x\mathbb{KL}(p(y|\mathbf{x})|p(y))) \\
    FID &= \lvert\lvert{m_1 - m_2}\rvert\rvert^2_2 + \mathrm{Tr}(C_1+C_2-2(C_1C_2)^{\frac{1}{2}}).
\end{align} 
\indent Inception score~(IS)~\cite{inception_score} is a metric for both image quality and diversity, which is found correlating well with the human evaluation. 
The conditional label distribution $p(y|\mathbf{x})$ measures the quality of the generated images, and the images are high-quality when the distribution is with low entropy. The marginal distribution $p(y)$ measures the diversity, and the generated images are more diverse when the distribution is with high entropy. 
By jointly considering them two together, the KL-divergence of $p(y|\mathbf{x})$ and $p(y)$ provides the evaluation for both image quality and its diversity. When calculating IS in our experiments, we use the Inception-v3 model finetuned on CUB-200 or Oxford-102 dataset following ~\cite{stackgan++}. 
Fr\'echet inception distance~(FID)~\cite{fid} measures the distance between the generated and real data. 
Lower FID means higher similarity for the generated and real data distribution. 
To calculate the FID between the generated images and the real images, we use all speech labels in the testing set to generate a large number images~(\textit{e.g.} 30k for CUB-200, 11k for Oxford-102). When calculating FID in our experiments, we use the Inception-v3 model pretrained on the ImageNet dataset~\cite{deng2009imagenet} following~\cite{fid}.

\subsection{Experimental Results on Synthesized Data}
\begin{figure*}[ht!]
    \begin{center}
       \includegraphics[width=0.96\linewidth]{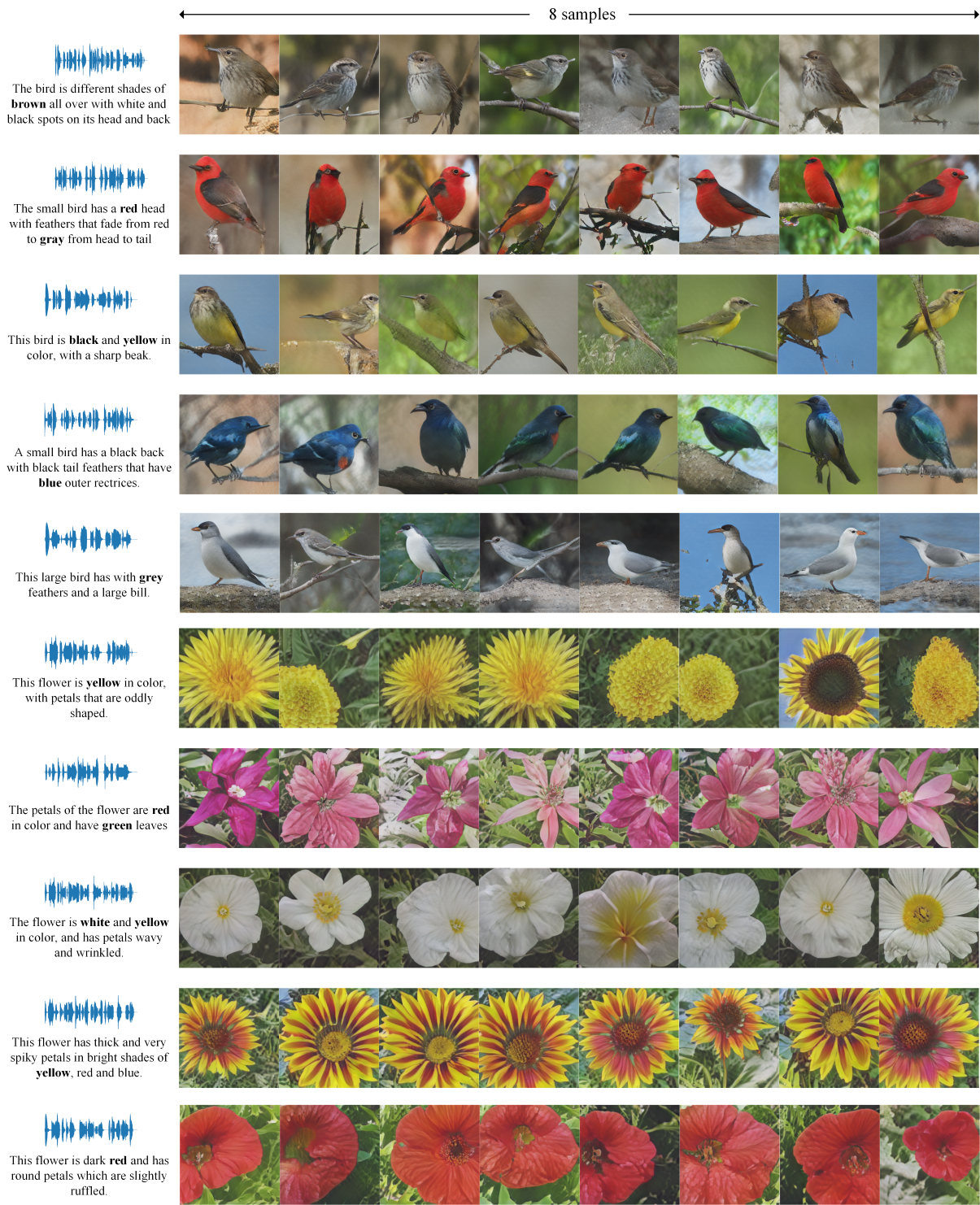}
    \end{center}
       \caption{Qualitive results for CUB-200~\cite{WahCUB_200_2011} testing set~(row 1$\sim$5) and Oxford-102~\cite{Flower} testing set~(row 6$\sim$10)~(synthesized speech data). Left: the waveforms of the input speech descriptions and its transcription results. Right: 8 images with a resolution of $256\times256$ synthesized by our model conditioned on the left input speech description and a random noise vector. \textbf{Note that the text is shown only for readability, and it is not used in our model.} Some interesting conclusions can be drawn from the figure: (1) the color of the birds/flowers are consistent with the input speech descriptions and they do not change with the input noise. (2) when fixing the speech descriptions, the backgrounds, gestures~(for birds), even the categories~(such as row 6) of the generated images change along with the input noise.}
    \label{fig:qualitative_result}
    \end{figure*}

To demonstrate the effectiveness of our model, experiments are conducted on the synthesized speech data, including CUB-200~\cite{WahCUB_200_2011} and Oxford-102~\cite{Flower} dataset. The synthesized speech data are continuous and unaligned, although they are well-normalized and less noisy. The CUB-200 dataset contains 200 classes totally, where 150 classes are used to train our model and the rest are used as the testing set. The Oxford-102 dataset contains 102 classes totally, where 82 classes are used for training and 20 classes for testing. 
The statistical information for training/testing split of these two datasets is shown in Table~\ref{tab:dataset_info}. The splitting manner for the training set and testing set follows~\cite{reed2016learning}. The qualitative and quantitative results are illustrated in Fig.~\ref{fig:qualitative_result} and Table~\ref{tab:IS_FID}, respectively. Some conclusions can be drawn from the results:

\begin{table}
  \begin{center}
  \begin{tabular}{l|c|c|c|c}
  \hline 
  \multicolumn{2}{c|}{Dataset} & Training set & Testing set & Total\\
  \hline
  
  \hline 
  \multirow{2}{*}{CUB-200~\cite{WahCUB_200_2011}} & class & 150 & 50 & 200 \\
   &image & 8855 & 2933 & 11788 \\
  \hline
  \multirow{2}{*}{Oxford-102~\cite{Flower}}& class & 82 & 20 & 102 \\
   &image & 7034 & 1155 & 8189 \\
  \hline
  \end{tabular}
  \end{center}
  \caption{Training/testing set of CUB-200~\cite{WahCUB_200_2011} and Oxford-102~\cite{Flower} datasets.}
  \label{tab:dataset_info}
  \end{table}

\subsubsection{Our model can synthesize images semantically consistent with the input speech}
qualitative results on CUB-200 testing set and Oxford-102 testing set are shown in Fig.~\ref{fig:qualitative_result}. For each row, we show the waveform of the input speech description~(left) and 8 synthesized images~(right) with different input noises conditioned on the same input speech embedding. 
It is worth mentioning that the transcription results are shown only for readability, and they are not used in our model. The results show that our model can generate realistic images from the input speech description although the input speech is continuous and unaligned. Moreover, the visual information shown in the synthesized images is mostly accordant with the semantic information in the speech descriptions. 
The generated results from CUB-200 dataest~\cite{WahCUB_200_2011}~(row 1$\sim$5 in Fig.~\ref{fig:qualitative_result}) show that the input speech description controls the color of the generated bird's different parts, such the head, feather, tail, etc. In comparison, the background, gesture, even shape of the generated bird changes when the input noise is different. This means that our model has disentangled the bird's color from the background and gestures to some extent. 
Similar conclusion also can be drawn from the results of Oxford-102~\cite{Flower} dataset~(row 5$\sim$10 in Fig.~\ref{fig:qualitative_result}). The input speech description mainly controls the color of the generated flowers, and the background, size, even the categories of the generated flowers are different when the input noises change. This demonstrates that our model has learned the semantic information in the input speech descriptions to some extent and visualized the semantic information onto the images.

\subsubsection{Our model is better than the ``two-stage'' model with text}
To compare our ``one-stage'' method~(translating speech to images without text) with the ``two-stage'' method~(using text as middle representation), we train the ``two-stage'' model on CUB-200 and Oxford-102 datasets. In our experiments, we use a pretrained ASR model, DeepSpeech\footnote{https://github.com/mozilla/DeepSpeech}~\cite{deepspeech, deepspeech2}, to transcribe the speech into text and use the (text, image) paired data to train a text encoder and a generator following~\cite{stackgan++}. 
For fair comparison, the generator and the hyper-parameters for its training in the ``two-stage'' method are the same as our model's. The results of the ``two-stage'' method on CUB-200/Oxford-102 are shown in the 1st/6th row of Table~\ref{tab:IS_FID}, respectively. The ``two-stage'' model is slightly inferior to our ``one-stage'' model on both datasets. 
On the CUB-200 dataset~\cite{WahCUB_200_2011}, our ``one-stage'' model performs better by $1.52$ on FID although the IS score of the ``two-stage'' method is comparable to our model's. Similarly, on the Oxford-102 dataset~\cite{Flower}, our ``one-stage'' method surpasses the ``two-stage'' method on FID~($54.76~\text{vs}~57.73$), while IS scores of the both models are the same~($3.23~\text{vs}~3.23$). We think the reason why the ``two-stage'' method performs worse is the word errors in speech recognition. 

{
  It is worth mentioning that DeepSpeech might be not a strong baseline in our experiment although DeepSpeech has millions of parameters and is trained on a large dataset, because our speech data are synthesized while the training data of DeepSpeech are real speech data. The word error rates~(WERs) of DeepSpeech on our datasets are above $50\%$, showing that DeepSpeech does not perform well on our synthesized speech data. To address this problem, in the following, we compare our proposed model with text-to-image models, which can be viewed as ``two-stage'' frameworks with an ideal ASR model~(WER is $0\%$), to evaluate the performance of our model.}

\subsubsection{Our method is comparable to the text-to-image methods}
To compare with the upper bound of the ``two-stage'' method, we compare our method with the text-to-image methods, which can be seen as ``two-stage'' methods with a perfect ASR model. 
Two text-to-image methods~\cite{stackgan, stackgan++}, which use similar structures with our model, are used in the experiments, as shown in Table~\ref{tab:IS_FID}. On CUB-200~\cite{WahCUB_200_2011} dataset, our model is slightly inferior to StackGAN-v2~\cite{stackgan++} on FID~($18.37~\text{vs}~15.37$), but performs better on IS~($4.09~\text{vs}~4.04$). We think that the reason is that our model generates more diverse results due to the speech signal is higher-dimensional than text signals. Compared with StackGAN-v1~\cite{stackgan}, our model performs better with a large gain because we use a stronger generator. 
On the Oxford-102 dataset~\cite{Flower}, our proposed model surpasses the StackGAN v1 on both FID~($54.76~\text{vs}~55.28$) and IS~($3.23~\text{vs}~3.20$), although it only surpasses the StackGAN v2 slightly.

This result has demonstrated that our model performs closely with the upper bound of the ``two-stage'' models, and our speech encoder has extracted the semantic information in the input speech descriptions into the embedding feature, which is subsequently used as the condition in the generator, just as the text-to-image model in~\cite{stackgan++}.

\subsubsection{Our teacher-student learning method is better than the classifier-based method}
To compare our teacher-student learning method with previous classifier-based methods~\cite{chen2017deep}, we train our speech encoder via the classified-based method rather than the teacher-student learning. Specifically, a classifier layer is added after the speech encoder and the speech encoder is trained with the cross entropy loss. To compare fairly, the classifier-based model uses the same feature extractor structure as our model's. In particular, we use the speech encoder in our model as a feature extractor and add a linear layer as the classifier, then train this classifier-based model on the training set. When testing, only the feature extractor is used. By this way, the parameters size and the hyperparameters for classified-based method and our proposed method are the same. The only difference is the training method. 
The results are illustrated in the 2nd and 7th row of Table~\ref{tab:IS_FID}. On the CUB-200~\cite{WahCUB_200_2011} dataset, our model performs rather better than the classifier-based method on both IS~($4.09~\text{vs}~3.68$) and FID~($18.37~\text{vs}~43.76$), indicating that our model's better generalization ability when using the same structure and parameters. 
On the Oxford-102\cite{Flower} dataset, FID score~($54.76~\text{vs}~64.75$) shows our model is better, however, the IS score~($3.23~\text{vs}~3.30$) draws different conclusions. IS takes both the image quality and diversity into account, so these results indicate that our model generates data closer to the real data but less diverse when compared with the classifier-based method on the Oxford-102 dataset. 
Consequently, our proposed method shows better FID and comparable IS on both CUB-200~\cite{WahCUB_200_2011} and Oxford-102~\cite{Flower} dataset due to its teacher-student learning method when compared with the classifier-based method.

\begin{table}
  \begin{center}
  \begin{tabular}{l|c|c|c|c}
  \hline 
   Dataset & Method & Type & IS $\uparrow$ & FID $\downarrow$\\
  \hline
  \multirow{5}{*}{CUB-200~\cite{WahCUB_200_2011}} & two-stage & speech & $4.04 \pm .04$ & $20.85$ \\
   &classifier-based & speech & $3.68 \pm .04$ &  $43.76$ \\
   &Ours & speech & $\mathbf{4.09 \pm .04} $   &  $ \mathbf{18.37} $  \\
   \cline{2-5}
   &StackGAN-v1~\cite{stackgan} & text & $3.70 \pm .04$ & $51.89$ \\
   &StackGAN-v2~\cite{stackgan++} & text & $4.04 \pm .05 $ & $15.03$ \\
  
  \hline

\multirow{5}{*}{Oxford-102~\cite{Flower}}  & two-stage & speech & $3.23~\pm .06$ & $57.73$ \\
  & classifier-based & speech & $\mathbf{3.30 \pm .06}$ & $64.75$  \\
  & Ours & speech & $3.23 \pm .05 $   &  $ \mathbf{54.76} $  \\
  \cline{2-5}
   & StackGAN-v1\cite{stackgan} & text & $3.20 \pm .01$ & $55.28$ \\
   & StackGAN-v2\cite{stackgan++} & text & $3.26 \pm .01 $ & $48.68$ \\
  \hline
  \end{tabular}
  \end{center}
  \caption{quantitative results of our model on CUB-200~\cite{WahCUB_200_2011} dataset and Oxford-102~\cite{Flower} dataset. We compare our model with ``two-stage'' method, classifier-based method~\cite{chen2017deep}, and text-to-image models~\cite{stackgan, stackgan++}. As illustrated in the table, our ``one-stage'' model performs better than the ``two-stage'' model and the classified-based model, even comparably to the text-to-image models.}
  \label{tab:IS_FID}
  \end{table}


  \begin{figure}[ht!]
    \begin{center}
       \includegraphics[width=1.0\linewidth]{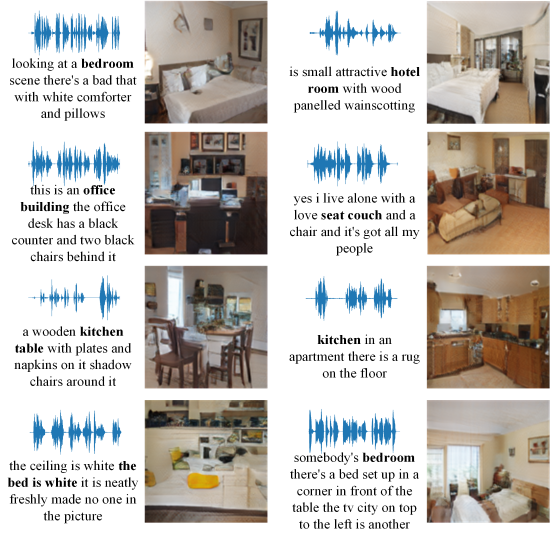}
    \end{center}
       \caption{Experimental results on Placces-Subset~(real speech data). Left: the waveforms of the input speech. Right: the synthesized images with a resolution of $128\times128$ conditioned on the left input speech. \textbf{Note that the text here is shown only for readability, and they are not used in our model.} The errors in the text come from the recognition error.}
    \label{fig:image_quality_places}
  \end{figure}

\subsection{Experimental Results on Real data}
In addition to the synthesized speech data, we also evaluate our model on the real speech data to assess the potential for real applications. Places Audio Captions dataset~\cite{harwath2016unsupervised,harwath2018jointly}, collected via Amazon Mechanical Turk~(AMT), is a real speech dataset for visual descriptions for Places 205 dataset~\cite{zhou2014learning}.
We use a subset of this dataset, which includes 13803 paired data with 7 classes\footnote{7 classes in Places 205: bedroom, dinette, dining room, home office, hotel room, kitchenette, living room}~(Places-Subset), to evaluate the robustness for the real data of our model. 
The testing set contains 2870 images, which are randomly selected from the dataset. It is difficult to generate the details for high-resolution images due to the dataset's diversity and the variance between different speakers, so we generate images with a resolution of $128\times128$. 
Some sampled results are shown in Fig.~\ref{fig:image_quality_places}. 
Although the details are not sharp due to the low resolution, we can also see that the color and the scene of the generated images are semantically consistent with the input speech descriptions, which means the model has captured the semantic information in the input speech descriptions to some extent. In addition to visualization examples, we also evaluate the result with objective matrics.
\begin{table}[t!]
  \begin{center}
  \begin{tabular}{r|c|c|c}
   \hline
   Dataset & Method & Type & FID$\downarrow$ \\
  \hline 
  \multirow{3}{*}{Places-Subset} 
  & two-stage & speech & 64.59 \\
  & classifier-based & speech & 232.39 \\
  & Ours & speech & 83.06 \\
  \hline
  \end{tabular}
  \end{center}
  \caption{Quantitative results of our model on Places-Subset dataset.}
  \label{tab:places}
\end{table}
Our model achieves $83.06$ for FID, as shown in Table~\ref{tab:places}~(IS is not used because no finetuned Inception model is available for Places-Subset). Our method is rather better than the classifier-based method~(83.06 vs 232.39), demonstrating the effectiveness of the teacher-student learning on the real data. Besides, different from the synthesized datasets, our model performs not as well as the ``two-stage'' method~(83.06 vs 64.59), because the real data are much challenging than the synthesized data to extract the speech semantic feature for our model.
These results are encouraging and show the potential for our model to the real scenario, such as human-computer interactions and computer-aided design.

\section{Ablation Study}\label{sec:ablation_study}
In this section, we conduct ablation study for our model to analyze the different components of the proposed model and the influence of some hyper-parameters as well as the embedding feature space. To compare fairly, we train the models from the scratch for the same iterations with the same hyperparameters except for the specified hyperparameter, such as the loss items for the speech encoders, or the scale of the synthesized images. Specifically, the speech encoders are trained for 100 epochs for both dataset, and the generators are trained for 220k/100k iterations for CUB-200~\cite{WahCUB_200_2011}/Oxford-102~\cite{Flower} dataset, respectively. The training iterations is set to avoid overfitting or model collapse.

\begin{figure*}[ht]
  \begin{center}
     \includegraphics[width=0.99\linewidth]{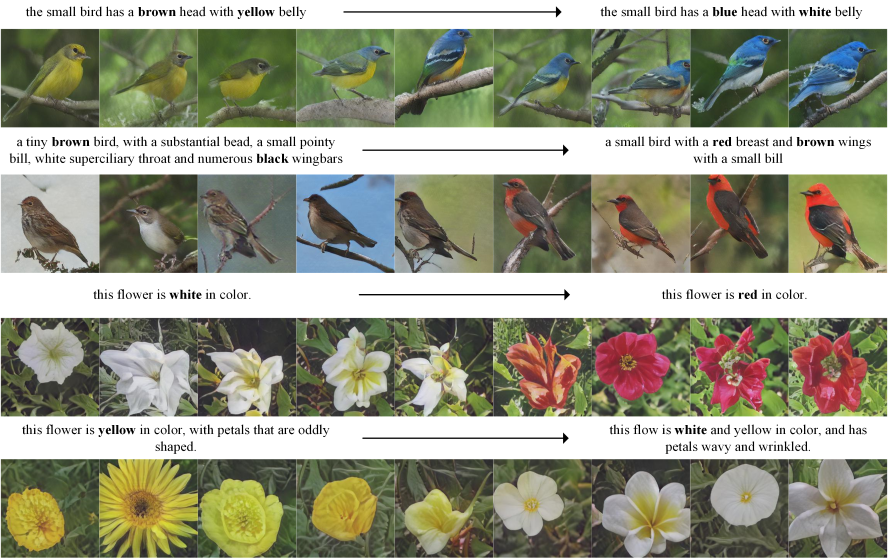}
  \end{center}
     \caption{Feature interpolation results on CUB-200~\cite{WahCUB_200_2011}~(row 1-2) and Oxford-102~\cite{Flower}~(row 3-4) dataset. \textbf{Note that the inputs are the speech descriptions without the text representations, the text is shown only for readability.} As illustrated in the figure, the generated images are semantically consistent with the input speech descriptions. Moreover, from left to right on each row, the color of the birds/flowers transits gradually due to the feature interpolation, which indicates that our model learns a linear semantic feature space.}
  \label{fig:feature_interpolation}
\end{figure*}

\begin{table}[t!]
  \begin{center}
  \begin{tabular}{l|c|c|c|c|c}
  \hline 
   Dataset & $\mathcal{L}_{L_{norm}}$ & $\mathcal{L}_{JEL}$ & $\mathcal{L}_{KDL}$ & IS$\uparrow$ & FID$\downarrow$\\
  \hline
  \multirow{3}{*}{CUB-200~\cite{WahCUB_200_2011}} & \checkmark &  &  & $3.71\pm .05$ & $27.45$ \\
   & \checkmark& \checkmark &  & $4.02\pm .04$ & $\mathbf{16.75}$ \\
   & \checkmark & \checkmark & \checkmark & $\mathbf{4.11 \pm .05} $   &  $ 18.70 $  \\
  \hline

\multirow{3}{*}{Oxford-102~\cite{Flower}} & \checkmark &  &  & $2.81\pm .04$ & $66.84$\\
& \checkmark& \checkmark & & $3.22\pm .03$ &  $58.19$  \\
& \checkmark & \checkmark & \checkmark & $\mathbf{3.23 \pm .05} $   &  $ \mathbf{57.11} $  \\
  \hline
  \end{tabular}
  \end{center}
  \caption{Ablation study results for different loss items of the speech encoder on CUB-200~\cite{WahCUB_200_2011} and Oxford-102~\cite{Flower} datasets.}
  \label{tab:abla_loss}
  \end{table}

\begin{table}[t!]
  \begin{center}
  \begin{tabular}{l|c|c|c}
  \hline 
   Dataset & Scale & IS $\uparrow$ & FID $\downarrow$\\
  \hline
  \multirow{3}{*}{CUB-200~\cite{WahCUB_200_2011}} & $64\times 64$& $3.45 \pm .04$ & $70.18$  \\
   & $128\times 128$& $3.98 \pm .04$ & $28.25$   \\
   & $256\times 256$ & $\mathbf{4.11 \pm .05}$   &  $ \mathbf{18.70} $  \\
  
  \hline

\multirow{3}{*}{Oxford-102~\cite{Flower}} & $64\times 64$ & $2.82 \pm .04$ & $70.95$ \\
& $128\times 128$ & $\mathbf{3.27 \pm .06}$  & $63.35$  \\
& $256\times 256$ & $3.23 \pm .05 $   &  $ \mathbf{57.11} $  \\
  \hline
  \end{tabular}
  \end{center}
  \caption{Ablation study results for different scales for the generator on CUB-200~\cite{WahCUB_200_2011} and Oxford-102~\cite{Flower} datasets.}
  \label{tab:abla_scale}
  \end{table}
\subsection{The Loss Function}
In the training of our speech encoder, teacher-student loss~(TSL) (Eq.~\ref{eq:se_loss}), including $\mathcal{L}_{L_{norm}}$, $\mathcal{L}_{JEL}$ and $\mathcal{L}_{KDL}$, is used to stabilize and accelerate the training. The experiments are conducted on CUB-200~\cite{WahCUB_200_2011} and Oxford-102~\cite{Flower} datasets to study how these items influence the final result. To compare fairly, the training hyper-parameters are all set the same except the loss items. 
The same structure and parameters are used in the generator training, and the generator is with three branches to generate images with a resolution of $256\times256$. The weight for $\mathcal{L}_{L_{norm}}/\mathcal{L}_{JEL}/\mathcal{L}_{KDL}$ is set as $5/1/1000$, respectively, to ensure the gradient to the speech embedding is within the similar scale~(following~\cite{speech2face}).
Table~\ref{tab:abla_loss} lists the results on the testing sets of CUB-200 dataset~\cite{WahCUB_200_2011} and Oxford-102 dataset~\cite{Flower}. When the only $\mathcal{L}_{L_{norm}}$ is used to train the speech encoder~(1st row and 4th row in Table~\ref{tab:abla_loss}), the model performs the worst on both datasets, demonstrating that $L_{norm}$ is not enough to obtain good performance. 
$\mathcal{L}_{JEL}$ improves the model with a big margin on both datasets~(2nd row and 5th row in Table~\ref{tab:abla_loss}), even achieves the best FID on CUB-200 dataset~(2nd row in Table~\ref{tab:abla_loss}). 
As a comparison, $\mathcal{L}_{KDL}$ only boosts the model slightly on Oxford-102 dataset~(6th row in Table~\ref{tab:abla_loss}) and even leads an FID drop on CUB-200 dataset~(3th row in Table~\ref{tab:abla_loss}). In general, both $\mathcal{L}_{JEL}$ and $\mathcal{L}_{KDL}$ can improve the model more or less.

\subsection{Different Scales}
Higher resolution provides us more details, making the generated images more realistic. However, higher resolution increases the complexity of the model, making the model unstable. The scale of the generated images affects the performance of our model in different aspects. So in this subsection, we conduct experiments to study the influences of different scales. 
Experiments are conducted on CUB-200 dataset~\cite{WahCUB_200_2011} and Oxford-102 dataset~\cite{Flower}. All the experiments use the same training hyperparameters and network structure, except for the resolution of the generated images and the discriminator number for the different scales. 
Specifically, all the experiments use the same pretrained speech encoder to extract embedding features of the speech descriptions.  The generator to synthesize images with a resolution of $64^2$/$128^2$/$256^2$ uses $1/2/3$ discriminators to model the data distribution, respectively. IS~\cite{inception_score} and FID~\cite{fid} are used to evaluate the performance of the generator.
As illustrated in Table~\ref{tab:abla_scale}, based on the tree-like structure~\cite{stackgan++}, the higher resolution our model synthesizes, the better performance we can achieve. The only exception is the model for the resolution of $128\times 128$ obtains the best IS on Oxford-102 dataset~(5th row in Table~\ref{tab:abla_scale}), however, it only performs better within $1.5\%$~($3.27~\text{vs}~3.23$) than the model for the resolution of $256\times256$. Taking the FID into account, we can still conclude that higher resolution leads to better performance. The similar conclusion is also drawn in the text-to-image translation~\cite{stackgan, stackgan++}.

\subsection{Feature Interpolation}
To study the feature space derived from our speech encoder further, we conduct experiments of feature interpolation on CUB-200~\cite{WahCUB_200_2011} and Oxford-102~\cite{Flower} datasets to show that the feature space learned by our speech encoder is a linear space to some extent. 
Specifically, given two embedding features $f_{s_1},f_{s_2}$, 9 features are sampled by combining $f_{s_1}$ and $f_{s_2}$ linearly: $f_{s_i} = \alpha_i f_{s_1} + (1-\alpha_i)f_{s_2}, \alpha=0,1/8,2/8,\ldots,1$. Then these embedding features are fed into the generator to study the semantic transition from $f_{s_1}$ to $f_{s_2}$. 
As illustrated in Fig.~\ref{fig:feature_interpolation}, in the first row, the color of the small bird's back shows a smoothing transition from brown to blue, while the color of belly changes from yellow to white, which exactly shows the semantic described in the input speech descriptions. 
In the second row, the breast of the small bird changes from brown to red smoothly, just as described in the input speech. The third row and the fourth row are from Oxford-102. Similar to the first two rows, most of the flowers are realistic and the color of the flower transits gradually from the left to the right, although there are some artifacts in some images. 
The results of feature interpolation verify that our model has learned a linear embedding feature space to some extent. 



\section{Conclusions}\label{sec:conclusion}
In this paper, we have described a new framework to translate the speech signals into the images without the help of middle text representation. We addressed this problem by extracting a low-dimensional embedding feature from the speech descriptions and synthesizing images from this feature via a stacked GAN. We have demonstrated that our proposed model can synthesize images semantically consistent with the input speech description on both synthesized and real data. Moreover, our model performed better than the ``two-stage'' method and the classifier-based method, even achieved comparable performance to the text-to-image models on the synthesized datasets. We believe that synthesizing images from speech signals without text is a new perspective to understand the semantic information in the speech signals and can open up new research directions.

\section{Acknowledgment}
This work was supported by the National Natural Science Foundation of China and Royal Society (61961130392), National Natural Science Foundation of China (61632001), which are gratefully acknowledged.

\ifCLASSOPTIONcaptionsoff
  \newpage
\fi



%
\bibliographystyle{IEEEtran}
\bibliography{refbib}

%



\begin{IEEEbiography}[{\includegraphics[width=1in,height=1.25in,clip,keepaspectratio]{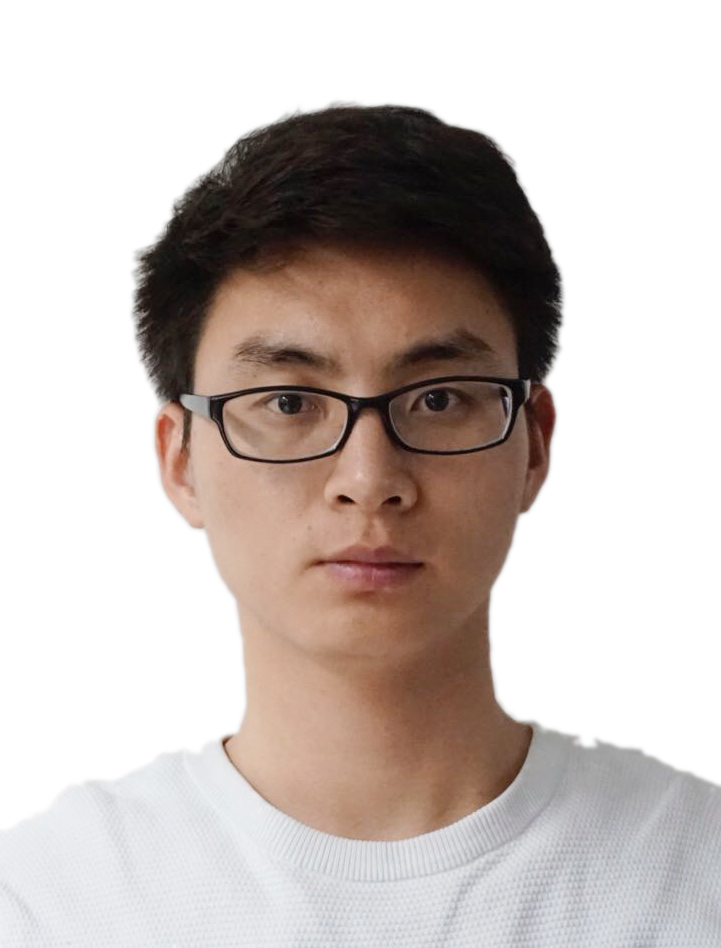}}]{Jiguo Li}
  received the B.S. degree in automation from Huazhong University of Science and Technology, Wuhan, China, in 2016. He is currently pursuing a Ph.D. degree with the Institute of Computing Technology, University of Chinese Academic of Science, Beijing, China. He is a visiting student with the Institute of Digital Media, Peking University from 2017. His research interests include cross-media intelligence and deep learning security.
  \end{IEEEbiography}

\begin{IEEEbiography}[{\includegraphics[width=1in,height=1.25in,clip,keepaspectratio]{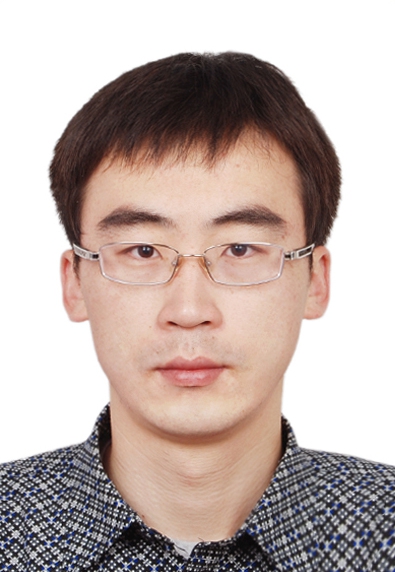}}]{Xinfeng Zhang}
(M'16) received the B.S. degree in computer science from the Hebei University of Technology, Tianjin, China, in 2007, and the Ph.D. degree in computer science from the Institute of Computing Technology, Chinese Academy of Sciences, Beijing, China, in 2014. He currently is an Assistant Professor with the School of Computer Science and Technology, University of Chinese Academy of Sciences. He authored more than 100 refereed journal/conference papers and received the Best Paper Award of IEEE Multimedia 2018, the Best Paper Award at the 2017 Pacific-Rim Conference on Multimedia (PCM) and the Best Student Paper Award in IEEE International Conference on Image Processing 2018. His research interests include video compression, image/video quality assessment, and image/video analysis.
\end{IEEEbiography}

\begin{IEEEbiography}[{\includegraphics[width=1in,height=1.25in,clip,keepaspectratio]{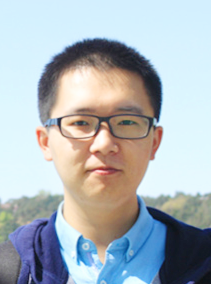}}]{Chuanmin Jia}
  received the B.E. degree in computer science from Beijing University of Posts and Telecommunications, Beijing, China, in 2015. He is currently pursuing a Ph.D. degree with the Department of Computer Science in Peking University, Beijing, China. He was a visiting student with Video Lab, New York University, NY, USA, in 2018. His research interests include visual data compression and machine learning.
 \end{IEEEbiography}
 \vskip 0pt plus -1fil
 \begin{IEEEbiography}[{\includegraphics[width=1in,height=1.25in,clip,keepaspectratio]{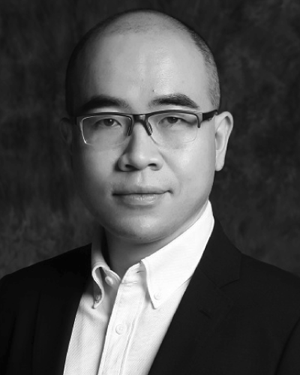}}]{Jizheng Xu}
  (M'07-SM'10) received the Ph.D. degree in electrical engineering from Shanghai Jiaotong University, China. In 2018, he joined Bytedance Inc. Before that, he was with Microsoft Research Asia as a Research Manager. He has authored or co-authored over 140 refereed conference and journal papers. He holds over 60 U.S. patents granted or pending in image and video coding. His research interests include image and visual signal representation, image/video compression and communication, computer vision, and deep learning. He is an active contributor to ISO/MPEG and ITU-T video coding standards. He has over 50 technical proposals adopted by international standards, including H.264/AVC, H.264/AVC scalable extension, High Efficiency Video Coding (HEVC), HEVC range extensions, HEVC screen content coding extensions and Versatile Video Coding. He chaired and co-chaired the Ad Hoc Group of exploration on wavelet video coding in MPEG and various technical ad hoc groups in JCT-VC, e.g., on screen content coding, on parsing robustness, and on lossless coding. He is an Associate Editor of IEEE Transactions on Circuits and Systems for Video Technology. He served as a Guest Editor for the Special Issue on Screen Content Video Coding and Applications for IEEE Journal on Emerging and Selected Topics in Circuits and Systems. He was a Co-Organizer and the Co-Chair of special sessions on scalable video coding, directional transform, and high-quality video coding at various conferences.
 \end{IEEEbiography}
 \vskip 0pt plus -1fil
 \begin{IEEEbiography}[{\includegraphics[width=1in,height=1.25in,clip,keepaspectratio]{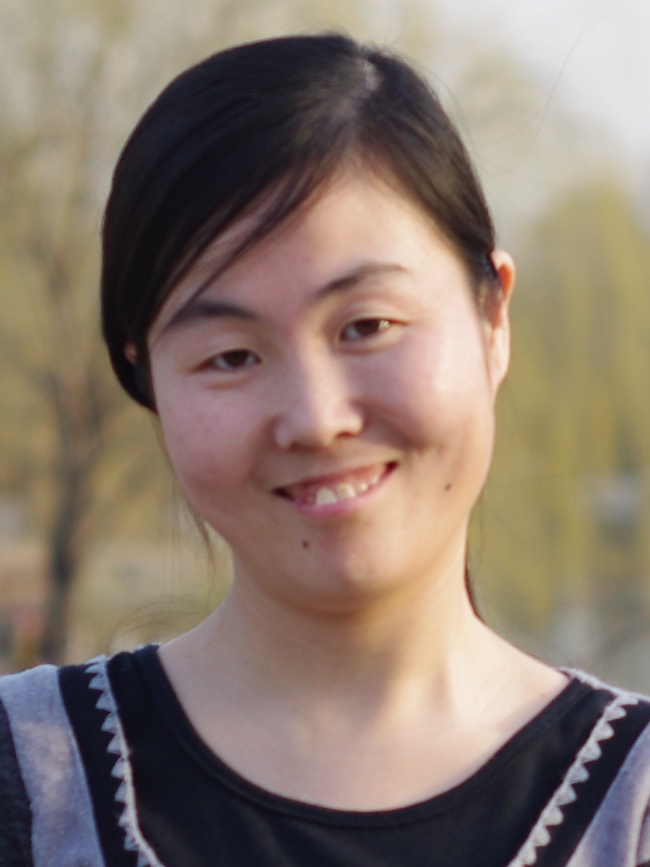}}]{Li Zhang} (M'07) received the B.S. degree in computer science from Dalian Maritime University, Dalian, China, in 2003, and the Ph.D. degree in computer science from the Institute of Computing Technology, Chinese Academy of Sciences, Beijing, China, in 2009. From 2009 to 2011, she held a post-doctoral position at the Institute of Digital Media, Peking University, Beijing. From 2011 to 2018, she was a Senior Staff Engineer, with the Multimedia R\&D and Standards Group, Qualcomm, Inc., San Diego, CA, USA. She is currently the Lead of the Video Coding Standard Team, Bytedance Inc., San Diego, CA, USA. Her research interests include 2D/3D image/video coding, video processing, and transmission. She was a Software Coordinator for Audio and Video Coding Standard (AVS) and the 3D extensions of High Efficiency Video Coding (HEVC). She has authored 300+ standardization contributions, 100+ granted US patents, 60+ technical articles in related book chapters, journals, and proceedings in image/video coding and video processing. She has been an active contributor to the Versatile Video Coding, Advanced AVS, the IEEE 1857, 3D Video (3DV) coding extensions of H.264/AVC and HEVC, and HEVC screen content coding extensions. During the development of those video coding standards, she co-chaired several ad hoc groups and core experiments. She has been appointed as an Editor of AVS, the Main Editor of the Software Test Model for 3DV Standards.
 \end{IEEEbiography}
 \vskip 0pt plus -1fil

 \begin{IEEEbiography}
  [{\includegraphics[width=1in,height=1.25in,clip,keepaspectratio]{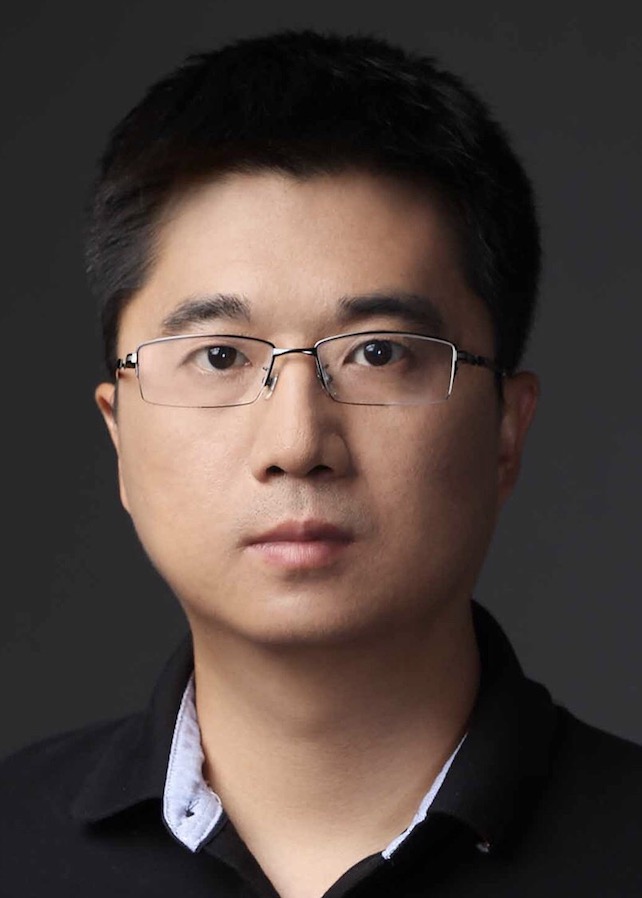}}]{Yue Wang}
received the B.S. degree in electronic engineering from Tsinghua University, Beijing, China, in 2006. He got Ph.D. degree with the Graduate University of the Chinese Academy of Sciences, Beijing, China in 2012. He did the postdoc research in Department of Computer Science in Peking University, Beijing, China between 2015 and  2017. He is now the director of multimedia foundation team in ByteDance Ltd since 2018. His current research interests include image and video coding, processing, and transmission technology.
\end{IEEEbiography}
\vskip 0pt plus -1fil

\begin{IEEEbiography}[{\includegraphics[width=1in,height=1.25in,clip,keepaspectratio]{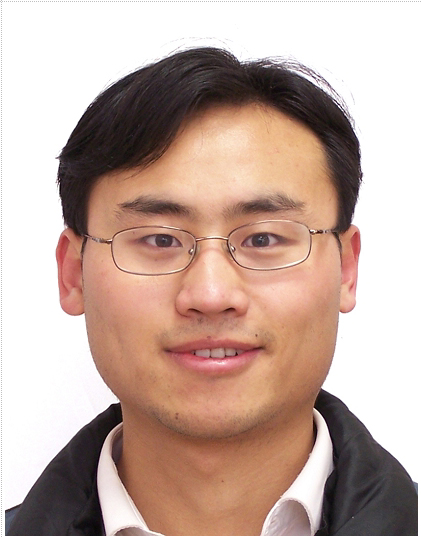}}]{Siwei Ma}
   (Member, IEEE) received the B.S. degree from Shandong Normal University, Jinan, China, in 1999 and the Ph.D. degree in computer science from the Institute of Computing Technology, Chinese Academy of Sciences, Beijing, China, in 2005. He held a Postdoctoral position with the University of Southern California, Los Angeles, CA, USA, from 2005 to 2007. He joined the School of Electronics Engineering and Computer Science, Institute of Digital Media, Peking University, Beijing, where he is currently a Professor. He has authored more than 300 technical articles in refereed journals and proceedings in image and video coding, video processing, video streaming, and transmission. He is an Associate Editor for the IEEE TRANSACTIONS ON CIRCUITS AND SYSTEMS FOR VIDEO TECHNOLOGY and the Journal of Visual Communication and Image Representation.
  \end{IEEEbiography}

  \vskip 0pt plus -1fil

    \begin{IEEEbiography}
      [{\includegraphics[width=1in,height=1.25in,clip,keepaspectratio]{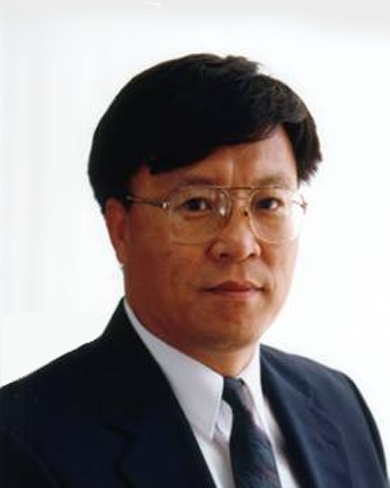}}]{Wen Gao}(M'92-SM'05-F'09) received the Ph.D. degree in electronics engineering from the Univer- sity of Tokyo, Tokyo, Japan, in 1991. He is currently a Professor of computer science with the School of Electronic Engineering and Computer Science, In- stitute of Digital Media, Peking University, Beijing, China. He is also the director of Peng Cheng Labo- ratory, Shenzhen, China, since 2018. Before joining Peking University, he was a Professor of computer science with the Harbin Institute of Technology, Harbin, China, from 1991 to 1995, and a Professor with the Institute of Computing Technology, Chinese Academy of Sciences, Beijing. He has authored extensively including five books and over 600 technical articles in refereed journals and conference proceedings in the areas of image processing, video coding and communication, pattern recognition, multimedia information retrieval, multimodal interfaces, and bioinformatics. He is a member of the China Engineering Academy. He has been the Chair of a number of prestigious international conferences on multimedia and video signal processing, such as the IEEE International Conference on Multimedia and Expo and ACM Multimedia, and served on the Advisory and Technical Committees of numerous professional organizations. He served or serves on the Editorial Board of several journals, such as the IEEE TRANSACTIONS ON CIRCUITS AND SYSTEMS FOR VIDEO TECHNOLOGY, the IEEE TRANSACTIONS ON MULTIMEDIA, the IEEE TRANSACTIONSON AUTONOMOUSMENTAL DEVELOPMENT, the EURASIP Journal of Image Communications, the Journal of Visual Communication, and Image Representation.
      \end{IEEEbiography}






\end{document}